
\input phyzzx
\input epsf
\overfullrule=0pt
\def\cp{{\it CP}}
\def\lfm{\smallskip\noindent\item}
\def\mmt{{\bf \bar m}}
\REF\krs{V. A. Kuzmin, V. A. Rubakov,
M. E. Shaposhnikov, {\it Phys. Lett.} B155 (1985) 36.}
\REF\farrar{ G.~R.~Farrar and M.~E.~Shaposhnikov, {\it Phys.\
   Rev.\ Lett.} 70 (1993) 2833,  hep-ph/9305274;  {\it Phys.\ Rev.} D 50 (1994)
   774, hep-ph/9305275, hep-ph/9406387.}
\REF\ghop{ M.~B.~Gavela, P.~Hern\'andez, J.~Orloff and O.~P\`ene,
   {\it Mod.\ Phys.\ Lett.} A 9 (1994) 795,  hep-ph/9406289;
{\it Nucl. \ Phys.}
 B430 (1994) 382,  hep-ph/9406289. }
\REF\hs{P.~Huet and E.~Sather, {\it Phys.\ Rev.} D51 (1995) 379,
hep-ph/9406345. }
\REF\review{For a review, see A. G. Cohen, D. B. Kaplan and
A. E. Nelson, {\it Ann. Rev. of Nucl. and Part. Sci.} 43 (1993) 27,
hep-ph/9302210.}
\REF\cknthin{ A.~G.~Cohen, D.~B.~Kaplan and A.~E.~Nelson,
{\it Phys. Lett.} 245 B (1990) 561;
\   {\it Nucl.\ Phys.} B 349 (1991) 727;
{\it Nucl.Phys.} B373 (1992) 453.}
\REF\lsvt{ L.~McLerran, M.~E.~Shaposhnikov, N.~Turok and
   M.~Voloshin, {\it Phys.\ Lett.} B 256 (1991) 451.}
\REF\cknthick{ A.G. Cohen, D.B. Kaplan, A.E. Nelson  {\it Phys. Lett.}
B263 (1991) 86.}
\REF\dhs{ M.~Dine, P.~Huet and R.~Singleton,~Jr, {\it  Nucl.Phys.} B375 (1992)
    625.}
\REF\dhss{ M.~Dine, P.~Huet, R.~Singleton,~Jr. and L.~Susskind,
   {\it Phys.\ Lett.} B 257 (1991) 351.}
\REF\cn{A. Cohen and A. Nelson,\ {\it Phys. \ Lett.} B297 (1992) 111. }
\REF\turokdiff{M. Joyce, T. Prokopec, N. Turok, PUPT-1436, (1994)
hep-ph/9401351;  PUPT-1494 (1994) hep-ph/9408339.}
\REF\ckndiffusion{A.~G.~Cohen, D.~B.~Kaplan and A.~E.~Nelson,
{\it Phys.Lett.} B336 (1994) 41, hep-ph/9406345.}
\REF\comelli{D.~ Comelli, M.~ Pietroni and A.~ Riotto, preprint DFPD-94-TH-39
(1994) hep-ph/9406369 .}
\REF\dhlll{ M.~ Dine,  P.~ Huet, R.~ G.~ Leigh, A.~
Linde and D.~ Linde, {\it Phys.\ Rev.} D 46 (1992) 550.}
\REF\mlt{ B.-H.~ Liu, L.~ McLerran, N.~ Turok, {\it  Phys.\ Rev.}
 D 46 (1992) 2668.}
\REF\hn{P.~Huet and A.~E.~Nelson, {\it in preparation}.}
\FIG\figone{(a) Contributions to $J_+$. (b) Contributions to $J_-$.}
\FIG\figtwo{(a) Amplitudes contributing to $T_L$. (b) Amplitudes
contributing to $R_L$.}
\FIG\figthree{(a) The form factor ${\cal I}$ is plotted
versus $\tau T$, for the case  $m \ll 0$,
$M_T = T/2$ (quark-gluon interactions).
The dotted curve results from
numerical integration of Eq. \formfactor; for values of
$\tau T > $ {\it a few},
it is well-approximated by its asymptotic form $1/2\sqrt{\tau T}$ (solid line).
(b) The dependence of ${\cal I}$ on the mass $\mmt$ is mild
in the range $\mmt \leq T$.}

\pubtype={UW/PT 95-03}
\date={June 1, 1995 (revised)}
\titlepage
\title{{\it CP} violation and Electroweak Baryogenesis in
                 Extensions of the Standard Model}
\author{Patrick Huet and Ann E. Nelson}
\address{Department of Physics \break
University of Washington \break
Box 351560
Seattle, WA 98195-1560}
\abstract{We develop a new and general
method to calculate the effects of \cp\ violation from extensions of the
standard model on
the mechanism of electroweak baryogenesis. We illustrate its applicability in
the framework of two-higgs doublet models.}

\endpage
\section{Introduction}

 It has recently been convincingly established that electroweak
baryogenesis\refmark\krs  \break due to
the mixing of three generations of quarks in the minimal standard model is
unable to account for today's baryon asymmetry.\refmark{\farrar,\ghop,\hs}
 This result is the direct consequence of a sharp conflict between the rapid
quark-gluon interactions and the far too slow processes of
quantum interference through
which the phase of the Kobayashi-Maskawa matrix can emerge into the physical
world.\refmark\hs

This result has long been anticipated and a plethora of
new sources of \cp\ violation have already been contemplated.
There is a clear contrast between these new sources of \cp\ violation and the
standard model CKM phase. The latter
is only physical because of the charge current interactions.
In contrast, in many models with new sources of \cp\ violation,
during the weak phase transition
some  mass matrix
 has a  space dependent phase   which cannot be removed    since making the
masses
real and diagonal at two adjacent points
$x$ and
$x+ dx$  requires, in a space-varying background,
two different unitary rotations $U_x$ and $U_{x+dx}$. The
relative rotation $ U_x^{-1}U_{x+dx}$ yields a
new interaction which can generate a \cp\ violating observable.
Because large physical \cp\ violating interference effects   can appear in the
phase
boundary where the particle masses are space dependent,    they play a dominant
role in the mechanism of electroweak baryogenesis.

At the electroweak phase transition,
non equilibrium \cp\ violating effects are largest inside the wall
of a bubble of broken phase expanding in a thermally equilibrated plasma.
Two apparently distinct mechanisms of electroweak baryogenesis have been
proposed in the
literature,\refmark\review
the so-called ``thin wall" and ``thick wall" scenarios. They are
characterized by the conditions $\ell/ L \gg 1$ and $\ell/L \ll 1$
respectively,
where $L$ and $\ell$ are the wall thickness and a typical mean free path of the
particles relevant to the scenario.

In the ``thin wall'' scenarios,
coefficients of reflection and transmission are computed.
For those species whose interactions with the wall take place through
a complex mass matrix, these coefficients assume different values for a
particle and
its \cp\ conjugate. The resulting asymmetry is then convoluted with
an incoming thermal flux to yield a \cp\ violating source which moves
along with the interface. This method has been applied to exotic fermions
with a majorana mass and to the top quark in two-higgs doublet
models.\refmark{\cknthin}

In the ``thick wall'' scenarios, thermally-averaged local operators have been
written which couple the baryon current, or a related current, to the
space-time derivative of the mass terms. These operators act as \cp\ violating
sources defined at every point of the interface. These are the scenarios
proposed in the context of the two
higgs\refmark{\lsvt,\cknthick,\dhs}
and supersymmetric models.\refmark{\dhss,\cn}

It has recently been understood that in both scenarios these sources
are to be inserted in a
set of coupled rate equations which allow for the \cp\ violating charges to be
transported  elsewhere in the plasma.
Transport greatly enhances the final baryon
asymmetry since anomalous electroweak baryon violating processes are
suppressed in the wall and in the broken phase
but are relatively rapid in the symmetric
phase.\refmark{\turokdiff,\ckndiffusion,\comelli}

This dichotomy in the formulation of electroweak baryogenesis mechanisms
obviously reflects limitations of the computational techniques and
does not do justice to the underlying physics of \cp\ violation.
 It is unsatisfactory for a few reasons.
\lfm{(a)} In practice, the thickness of the wall is neither very small nor very
large
compared to a mean free path. The ``thin wall'' approximation can  overestimate
the
magnitude of the actual baryon asymmetry produced while the ``thick wall''
approximation can underestimate it.
\lfm{(b)} An interpolation  between these two limits
is required as the \cp\ violating sources are the only inhomogeneous terms in
the rate equations so that the uncertainty in the baryon asymmetry is directly
proportional to the uncertainty in their determination.
\lfm{(c)} For any given energy, particles moving at
an
oblique enough angle relative to the wall are likely to be scattered while
inside the
wall. Hence the  integral over all particle momenta will include both particles
which
scatter inside the wall many times and those which do not scatter at all
and the ``thin wall" limit is never fully applicable.
\lfm{(d)} It has previously been argued
\refmark{\lsvt-\cn,\ckndiffusion}
that for a sufficiently thick wall it is a good approximation to compute the
effects
of the nonequilibrium \cp\ violation by adding time varying \cp\ and
{\it CPT} violating
terms to
an effective Hamiltonian which is treated as approximately spatially constant,
and by assuming that the local particle distributions relax towards thermal
equilibrium
with this Hamiltonian according to  some classical rate equations. However, the
plasma includes many particles with a small momentum perpendicular to the wall
and so with a long wavelength perpendicular to the wall. When this wavelength
is long compared with the particle mean free path,
a  ``classical'' treatment of the \cp\ violation is not adequate.

What follows is the description of a new method of calculation
which applies to all scenarios and  all values of $\ell/L$.
It reflects in the most direct way the  interplay between the coherent
phenomenon of \cp\ violation and the incoherent nature of the plasma
physics. This method can account for the generation of a \cp\
violating observable from mass matrices with
non-trivial space-dependence,
as well as from particle interactions.\foot{The latter mechanism
dominates when the
former one is not present. This is the case in the minimal standard
model
with \cp\ violation originating from the quark Yukawa couplings.\refmark\hs}
In its simplest form,
it easily reproduces the ``thin wall'' and ``thick wall''
calculations wih significant
improvements over earlier estimates. In its more general form, it
can incorporate  effects which arise from the
large diversity of scales present in a realistic plasma and can be the
basis for Monte-Carlo simulations.
For reasons of clarity, the method is best introduced with an
example: the two higgs doublet model. The
reader should bear in mind that it applies to other theories as well.

\section{Construction of the \cp\ violating sources}

Let us consider a set of particles with (not necessarily diagonal) mass matrix
$M(z)$
and moving, in the rest frame of the wall, with energy-momentum $E,{\vec
k}$.
At their last scattering point $z_o$, these particles emerge from a thermal
ensemble, propagate freely during a mean free time $\tau\sim\ell$, then
rescatter and
return to the local thermal ensemble in the plane $z_o+  \tau v$,
$v$ being
the velocity perpendicular to the wall, $ k_\perp /E$.
During the time $\tau$, these
particles evolve  according to a set of
Klein-Gordon, Dirac or Majorana equations coupled through the mass matrix
$M(z)$. It is in the course of this evolution that \cp\ violation
affects  the distribution of these particles.
Initially, at $z_o$, the contribution of these particles to any given charge
cancel exactly the contribution of their antiparticles:
$\langle Q \rangle = Tr [ {\hat Q}\,-\,{\hat Q}]\,=\,0$; here, $\hat Q$ is the
charge
operator and the trace is taken over flavors as well as particle distributions.
However,
after evolving a time $\tau$ across the \cp\ violating space-dependent
background, this cancellation no longer takes place
for those charges which are explicitly
violated by the mass matrix $M(z)$. At the subsequent scattering
point $z_o+\tau v$, these charges become
$\langle Q \rangle  = Tr[  A^\dagger {\hat Q} A
\,-\,{\overline A}^\dagger {\hat Q}
{\overline A}]$ and
assume a non-zero value, as
$A$, the evolution operator over the distance $\tau v$, is distinct
from its \cp\ conjugate $\overline A$.

To be specific, let us define $J_\pm$, the average current resulting
from particles moving toward positive(negative) $z$ between $z_o$ and
$z_o+ \Delta $, $\, \Delta = \tau v$. The current $J_+$ receives
contributions from either
particles originating from the thermal ensemble at point $z_o$, moving
with a positive velocity and being transmitted at $z_o+ \Delta$, or from
particles originating at $z_o+ \Delta$, moving with velocity $-v$ and
being reflected back towards $z_o+\Delta$(Fig. 1a). A similar definition
exists for $J_-$(Fig. 1b). $J_\pm$ are \cp\ violating currents
which are associated with each layer of thickness $\Delta$ moving along with
the wall. Once boosted to the plasma frame, these currents provide
 \cp\ violating sources, which fuel
electroweak baryogenesis.

The calculation
of these currents is facilitated by the use of $CPT$ symmetry and
unitarity.
$CPT$ symmetry identifies the amplitude for a
particle to be transmitted from the left with the amplitude
for its \cp\ conjugate to be transmitted from the right, while unitarity
relates transmission to reflection amplitudes. Instead of writing a
cumbersome but general formula for these currents, let us work them
out for a specific situation:
 a single fermion  with a Dirac  mass $M(z) = m(z) e^{i
\theta(z)}$.
It could be a top quark having its mass generated from a two
higgs-doublet lagrangian with an explicit \cp\ violating term in the higgs
potential, in which case, $\tau$ is the mean free
time for quark-gluon scatterings. As for the current,
we choose the axial current\foot{We choose this current because in the two
Higgs
model a combination of axial top number and Higgs number diffuses efficiently
into
the symmetric phase and is approximately conserved
by scattering in the symmetric phase.\refmark{\cknthick,\ckndiffusion}}. For
this
situation, the four-vectors
$J_\pm$ take the form
$$\eqalign{
J_+( z_o) = & \int_{\tilde v>0} {d^3{\vec k} \over (2
\pi)^3}
          \biggl\langle\bigl[  n(E,v)   -
                 n(E,-{\tilde v})\bigr]
                        \, Q(z_o,{\vec k},\tau)\biggr\rangle_{ z_o}\,\,
             (1,0,0,  {\tilde v}) \cr
J_-( z_o) = & \int_{\tilde v>0} {d^3{\vec k} \over (2 \pi)^3}
          \biggl\langle  \bigl[  n(E,v)   -
                 n(E,-{\tilde v})\bigr]
                        \, Q(z_o,{\vec k},\tau)\biggr\rangle_{ z_o}\,\,
(1,0,0, - v)\, . \cr}
\eqn\jpm$$

In this expression, $v$, $= k_\perp/E$, is the velocity perpendicular to
the wall at point $z_o$, ${\tilde v}$ is the velocity a distance
$\Delta$ away,
${\tilde v}^2 = v^2 + (m^2(z_o) - m^2(z_o + \Delta) )/E^2$ and
$n (E,v)$ is the Fermi-Dirac distribution $n_f$ boosted to the rest frame of
the wall, $=[\exp[\gamma_W E(1-v v_W)]+1]^{-1}$.
$Q(z_o,{\vec k},\tau)$ is the charge asymmetry which results
from the propagation of particles
of momentum ${\vec k}$ in the interval $[z_o,z_o+\Delta]$.
In our specific example, $Q$ is the chiral charge and is
given by
$$
Q_A(z_o,{\vec k},\tau) =
|T_L|^2 - |T_R|^2 -|T_{\overline L}|^2 + |T_{\overline R}|^2,
\eqn\chiralcharge
$$
where
$T_L$ is the amplitude for a left-handed spinor to propagate over the
distance $\Delta$,
$ T_R = T_L (M \rightarrow -M^\dagger)$ and $T_{\overline L} =
 T_L(M \rightarrow M^*)$. Finally, the brackets $\langle \ldots\rangle_{z_o}$
in Eq. \jpm,
average the location of point $z_o$ within a given layer of thickness
$\Delta$ as scattering occurs anywhere within a layer.

The standard ``thin wall'' and ``thick wall''
situations are obtained in taking  $\tau/L$  to $\infty$ and
$0$, respectively.
In the ``thin wall'' limit, $\,\tau/L \to \infty$,
the amplitudes $T$ become the usual transmission
coefficients and our expressions \jpm\ for $J_\pm$, match trivially earlier
calculations of scattering of particles off a sharp interface.
In the ``thick wall''
limit, $\,\tau/L \to 0$,
the currents $J_\pm$  yield, after a boost to the thermal frame,
a locally defined space-time dependent
source density $S(x,t)$
which generalizes, and gives a precise meaning to, the local \cp\ violating
operators already considered in the literature.
 In our example, the
source per unit volume per unit of time, located at
a point $x$ fixed in the plasma, at any given time $t$, is, to first order
in $v_W$ and $\tau/L$,

$$S_A(x,t) =  {-\gamma_w v_w \over 2 \pi^2}
            \int_0^1 dv
                \int_{z_o-\tau v/2}^{z_o+\tau v/2} {dz \over \tau v}
              \int_{\gamma m(z)}^\infty dE E^3 \ {dn_f \over dE}
          (2 v) \ {Q_A(z,{\vec k},\tau)
               \over \tau} \bigg|_{z_0=\gamma_w (x-v_w t)}.\eqn\source
$$
In order to obtain an explicit form for the source $S_A(x,t)$,
we need to compute the \cp\ violating charge $Q_A$
and perform the  integration in Eq. \source .

\section{Computation of the charge asymmetry $Q$}

In general,$Q(z_o,{\vec k},\tau)$ is a charge asymmetry produced
by particles moving with momentum $\vec k$ between the planes $z_o$ and
$z_o + \tau v$.  Its calculation may require a
different technique depending on
the relative values of the time scales involved and on the choice
of the charge.
The physics of the generation of a \cp\ violating observable,
is the physics of quantum interference. It is most easily dealt with
by treating the mass $M(z)$ as a small perturbation ($M(z) < T$). Using
techniques developed in Ref. \hs, one finds, for the transmitted amplitude,
$$
T_L(z_o,\tau) = e^{i \Delta k_\perp}\ \ \bigl[\,
1 \ - \ \int_{z_o}^{z_o+\Delta} dz_1 \int _{z_o}^{z_1} dz_2
\ e^{i2 k_\perp (z_1-z_2)} \  M (z_2) M^\dagger(z_1) \ + \
 {\cal O}(M/ k_\perp)^4\,\bigr].\eqn\tl
$$
 This expression has a straightforward diagrammatic formulation
presented in Fig. 2a. A similar expansion can be written for the reflection
amplitude $R_L$ (Fig. 2b). The various terms in the sum
correspond to various paths with
different \cp\ odd and \cp\ even phases. Only interference
between these paths contributes to a \cp\ violating
physical observable such as
$Q_A(z_o,{\vec k},\tau)$; one finds
$$
Q_A(z_o,{\vec k},\tau) =
8 \int_{z_o}^{z_o+\Delta} dz_1 \int _{z}^{z_1} dz_2 \
\sin 2 k_\perp (z_1-z_2)\ \  {\rm Im} [M (z_2) M^\dagger(z_1)]\ + \
 {\cal O}(M/k_\perp)^4 .\eqn\qtwo
$$
This expression is valid for any wall size and shape, and
generalizes easily
to the case where  many flavors mix. Given a wall
profile, the integrals can be evaluated and Eq. \qtwo\ can be
inserted in formulas
\jpm\ to provide an explicit form for the currents $J_\pm$.

It is   simplest to work out the case of a
very thick wall $L\gg\tau$. Using the derivative expansion
$M(z_i) = M(z_o)+(z_o-z_i)\partial_z M (z_o)$, one finds
$$
\eqalign{
Q_A(z_o,{\vec k},\tau) =& -4 f(k_\perp\Delta) /  k_\perp^3  \
{\rm Im} [M^\dagger\,\partial_z M]_{z_o} \cr
       =& -4 f(k_\perp\Delta) / k_\perp^3 \
m^2 \,\,\partial_z \theta|_{z_o} \qquad {\rm with} \qquad  f(\xi) =
 \sin \xi\, \bigl( \sin \xi - \xi \, \cos \xi \bigr). \cr} \eqn\qtwothick
$$
Inserting this latter expression into our formula \source\ for the source
density $S_A$ yields
$$
S_A(x,t)\,= \,-T \, \gamma_w v_w  \,
m^2 \partial_{z_o} \theta \big|_{z_o=\gamma_w (x-v_w t)} \,
 \times \ {2 \over \pi^2}\,
{\cal I}(\tau , m, T)+{\cal O}\biggl(v_w^2, (m/T)^4, (\tau/L)^2
 \biggr).
\eqn\axialsource
$$
${\cal I}(\tau , m, T)$ is a form factor whose general form is

$$
\eqalign{
{\cal I}(\tau, m, T) \simeq&{1 \over \sqrt{\tau T}} \
\int_{\mmt \over T}^\infty dy \ \sqrt{y} \  {e^y \over  (e^y + 1)^2}
\int_0^{\tau T (y- ({\mmt\over T})^2 {1\over y})}  {dt \over t^{3/2}} \ f(t)
\cr &{\rm with} \qquad \mmt^2 = m^2+ M_T^2\, . \cr }
 \eqn\formfactor
$$
We have included thermal
corrections, $M_T$, in the mass dependence of  Eq. \formfactor to take
 into account modifications of
particle dispersion relations from   scattering.
The effects of scattering  on  particle propagation
can be accounted for by
substituting quasiparticles for particles, in which case, $\tau$
is to be replaced with $1 /2\gamma$,
where $\gamma$ is the width of the quasiparticle. Correspondingly,
the dispersion relation is to be modified
to incorporate self-energy thermal corrections.
In the particular case of quarks scattering off gluons,
the width $\gamma$ is $ \simeq g^2_s T/3$, while the main thermal
corrections amount
to the shift $E^2 \to E^2 + M^2_T $, with
$M_T^2=\, g^2_s T^2/6 \simeq \,T^2/4$.
\foot{ A systematic method which accounts for
the thermal structure of a quasiparticle in the interference mechanism is
given in Ref. \hs.}

The form factor ${\cal I}$ is plotted in Fig. 3.
${\cal I}$ vanishes as $\tau \to 0$, it peaks at $\tau T
\sim 1$ and is well approximated by $\sim 1/\sqrt{\tau T} $ in the range
$\tau T > 5$. The interpretation of this behavior is
straightforward. As explained earlier, constructive interference
is maximal for particles whose transverse
Compton wavelength $k_\perp^{-1}$ is of the order of $\tau$,
that explains the peak at
$\tau T \sim 1$. As $\tau T$
increases,
fast oscillations along the distinct paths tend to cancel
against
each other and the resulting asymmetry drops; as a matter of fact,
in the extreme limit
$\tau T \gg 1$, the propagation is semi-classical and the asymmetry
vanishes as it should.\foot{However, a semi-classical treatment alone
might miss the important contributions of long-wavelength
particles moving at large angles in respect
to the wall motion; without their contribution, the asymmetry would
fall off as fast as $1/\tau$. }

In the opposite limit, as $\tau T \to 0$, the
asymmetry vanishes as the quantum coherence required
is washed away by the rapid plasma interactions.
Fig. 3b demonstrates the mild dependence of $\cal I$ on $\mmt$.

For the sake of comparison, we present an approximate form of
Eq. \axialsource, valid for $m \ll T$,
$$
S_A(x,t) \, \simeq \,  -{1 \over \pi^2} \, {1\over \sqrt{\tau T}}\,
\gamma_w v_w T\ m^2
\partial_z \theta  \qquad {\rm for} \qquad \tau T \geq 5
\eqn\sourcethicka
$$
$$
S_A(x,t) \, \simeq \, \,\, -{1\over 2 \pi^2}\,\,\,\,\,
\gamma_w v_w T\ m^2
\partial_z \theta  \qquad {\rm for} \qquad \tau T \sim 1-2.
\eqn\sourcethickb
$$

\section{Application to two-higgs models}

The formula we derived for the chiral source $S_A(x,t)$ can be directly
applied to the top quark propagating in the thick wall of a bubble
produced at the electroweak phase transition in two-higgs models.
Here the mean free path is dominated by
gluon scattering $\tau \sim  3/(2 g_s^2T)\sim (1-2)/T$ and is
typically smaller
than the estimated thickness of the wall:
$\tau /L \sim 0.01-0.1$.\refmark\dhlll
The wall velocity is $\gamma_w v_w < 1$.\refmark{\dhlll,\mlt} Finally,
The mass of the top quark $m_t(z)$ is
$ Y_t \phi \leq Y_t T \simeq T$ while $M_T \simeq T/2$, hence,
$\mmt = \sqrt{m^2_t + M_T^2} \leq 1.1 T$ and the assumptions
under which we derived Eq. \sourcethickb\ are approximately fulfilled.
We find
$$
S_A(x,t) \, \simeq \,  -{ N_c\over 2 \pi^2}\,
\gamma_w v_w T\ m^2
\partial_z \theta \ + \ {\cal O}\biggl(v_w^2, (\tau/L)^2, (m_t/T)^4\biggr),
\eqn\sourcethickb
$$
where the number of colors $N_c=3$.

Recent work on the source terms for axial top number in the two Higgs model
 in the thick wall case, have treated the fermion interaction terms with the
background Higgs field as a \cp\ violating contribution to a classical
Hamiltonian in computing the \cp-violating perturbation to particle
distributions. Ref. \cknthick\ pointed out that these
interactions split the energy levels of particles and their
CP conjugates in a
way reminiscent of a chemical potential.
These classical treatments obscure the origin of the \cp\ violating
effect as resulting
from  quantum
interference.
However, these methods, if implemented properly,
should provide reasonable approximations to our formulae for
those particles whose  wavelength $1/k_\perp$
is short compared with $v\tau$.
As an illustration, Ref. \ckndiffusion\ found a source term
$S_A(x,t) =   1/3 \, T\  v_w m^2 \partial_z \theta$. These authors
did not account  for the   quark-gluon interactions
in the rate for incoherent axial top number violation
( a factor $1/\tau$) and
the three dimensional phase space (a factor of $\sim 9\zeta(3)/\pi^2$), factors
which are
numerically unimportant but which are needed for theoretical
self-consistency.
 Even
after
including these effects, our formulae do not agree for large $\tau$ because our
integral~\source\ is dominated by particles with long wavelengths in the
direction
perpendicular to the wall, for which a classical approximation is not adequate.
Numerically, for
$\tau T\sim 2$, our answer approximately agrees in magnitude.

\section{Conclusion}

In conclusion, we have introduced a new method to compute \cp\ violating
sources resulting from particle interaction with an expanding
bubble during a first-order electroweak phase transition.
\pointbegin The method refers to explicit physical processes in the plasma.
In particular, it does not make use of thermally averaged operators,   or
effective
chemical potentials whose connection to the  microphysics is indirect, as they
 do not
vanish as the mass $m$ and/or the mean free time $\tau$  vanish, and whose
applicability is restricted to the range
$L \gg\tau\gg 1/T$.
\point The method makes explicit the quantum physics of \cp\ violation
and its suppression resulting from thermal effects.
\point  A major advantage of our formulation is that it
easily applies to charges generated
by flavor mixing through arbitrary large mass matrices. In particular,
it can be applied to cases,
such as the supersymmetric standard model\refmark\hn for which there
is no known semi-classical approximation.
\point It is valid for all wall shapes and sizes
as well as for arbitrary
particle species and interactions.
\point Finally, it incorporates \cp\ violation which originates from
particle interactions as well as from
non-trivial space-time mass dependences.
In particular, it generalizes and agrees with the decoherence
arguments invoked to rule out electroweak baryogenesis from \cp\
violation in the quark Yukawa couplings, as given in Ref. \hs.

\ack
This work was supported in part by the DOE under contract
\#DE-FG06-91-ER40614. The work of A. N. was supported in part by a
fellowship  from the Sloan Foundation.

\figout
\refout
\vfill\eject

\epsfbox{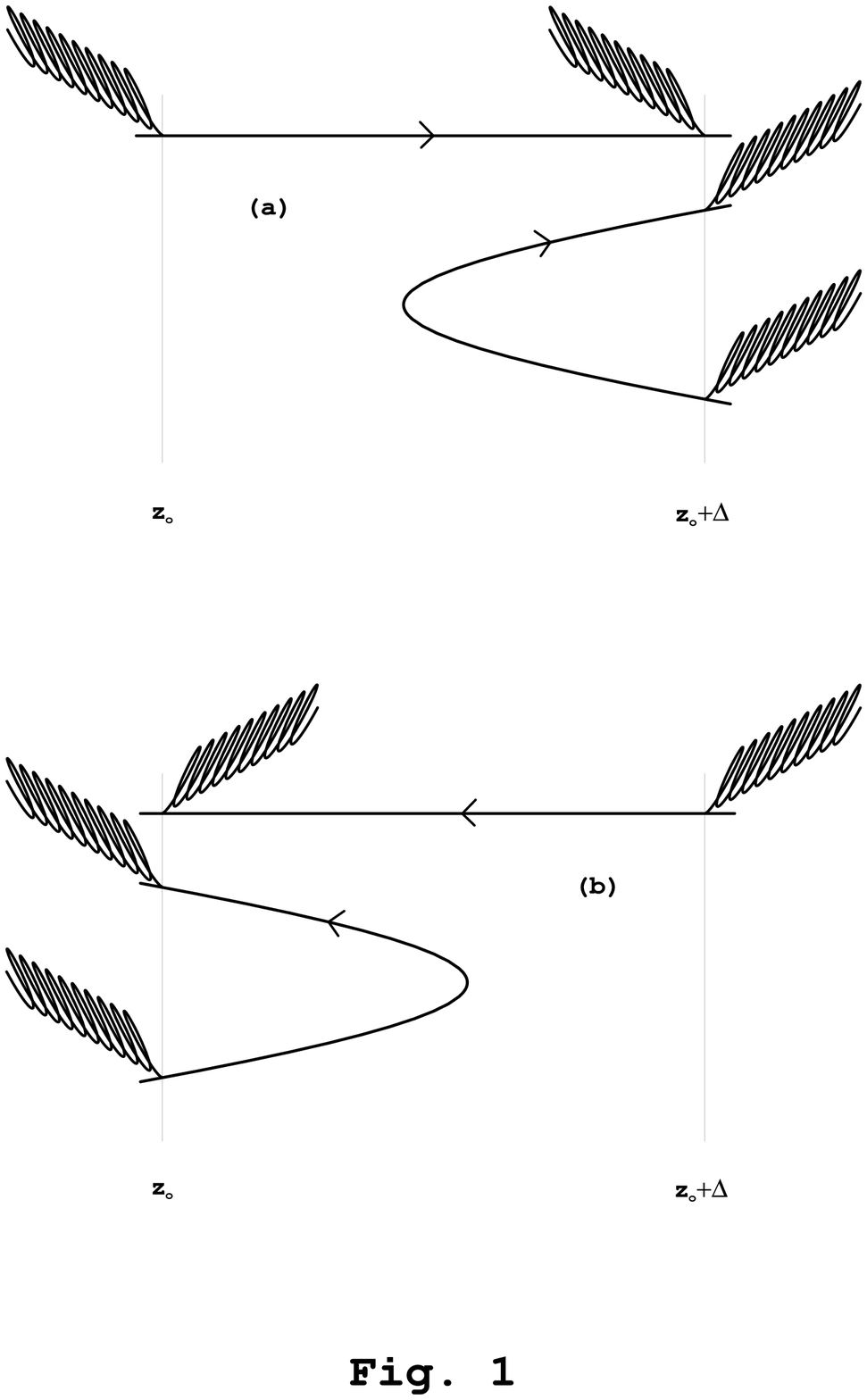}
\epsfbox{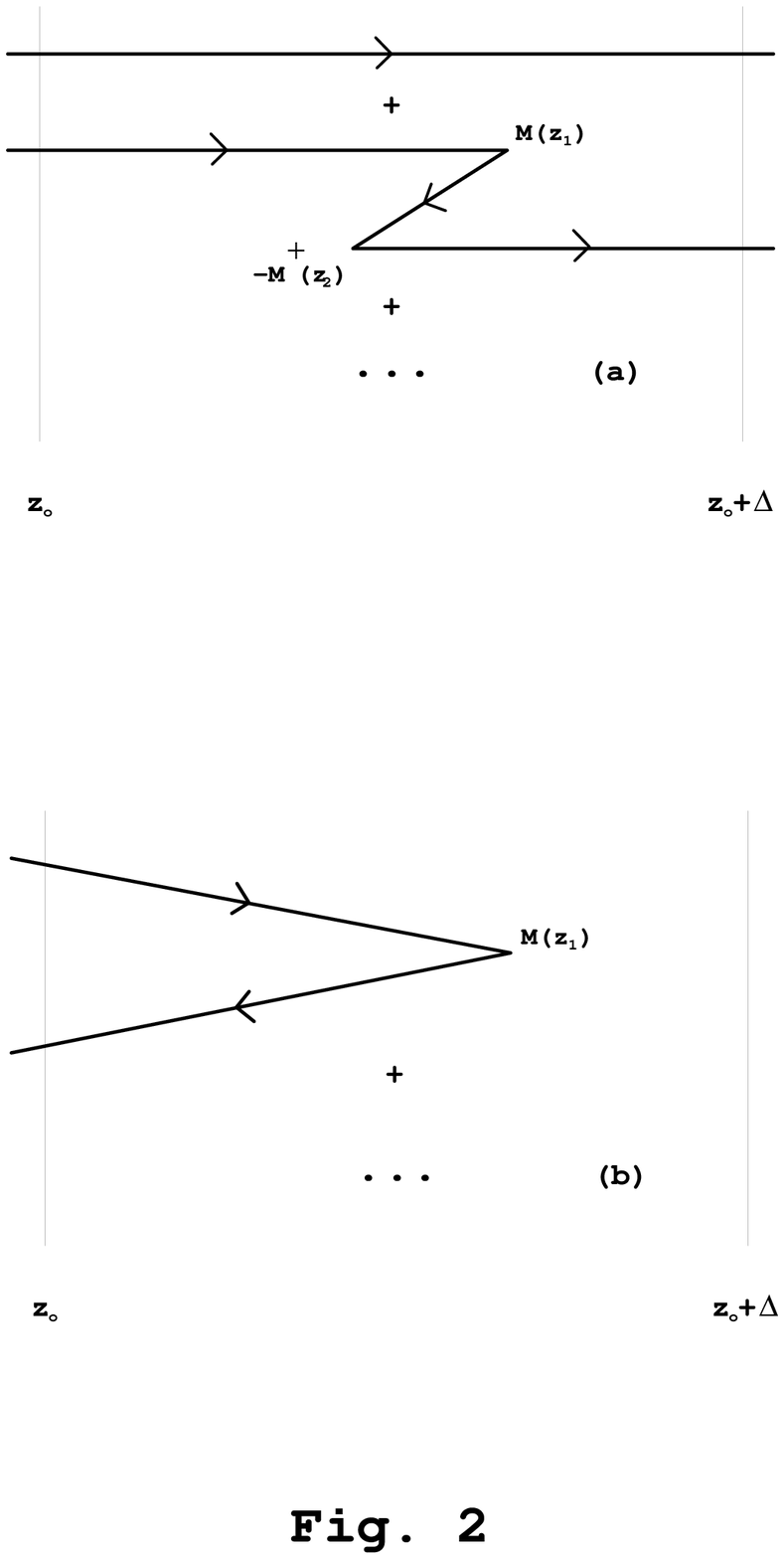}
\epsfbox{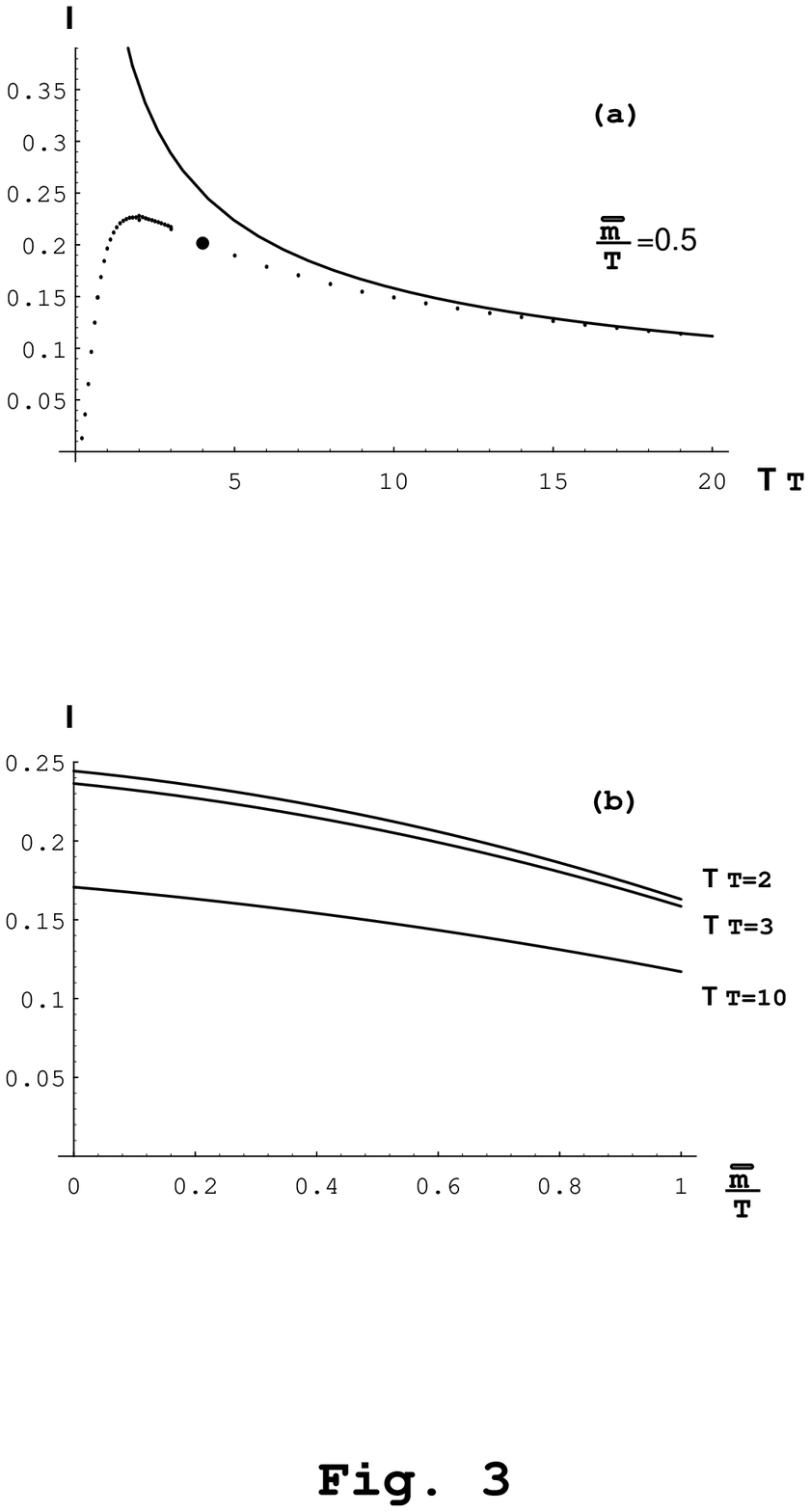}

\bye
\end